\def\year{2018}\relax
\documentclass[letterpaper]{article} 
\usepackage{alec18}  
\usepackage{times}  
\usepackage{helvet}  
\usepackage{courier}  
\usepackage{url}  
\usepackage{graphicx}  
\begin{document}
 
%
\title{Coordination-driven learning in multi-agent problem spaces}
\author{Sean L. Barton, Nicholas R. Waytowich, and Derrik E. Asher\\
U.S. Army Research Laboratory, Aberdeen Proving Ground\\
Aberdeen, Maryland 21005\\
}
\maketitle
\begin{abstract}

We discuss the role of coordination as a direct learning objective in multi-agent reinforcement learning (MARL) domains. To this end, we present a novel means of quantifying coordination in multi-agent systems, and discuss the implications of using such a measure to optimize coordinated agent policies. This concept has important implications for adversary-aware RL, which we take to be a sub-domain of multi-agent learning.

\end{abstract}

\section{Introduction}
Modern reinforcement learning (RL) has demonstrated a number of striking achievements in the realm of intelligent behavior by leveraging the power of deep neural networks \cite{Mnih2015aa}. However, like any deep-learning system, RL agents are vulnerable to adversarial attacks that seek to undermine their learned behaviors \cite{Huang2017aa}. In order for RL agents to function effectively along side humans in real-world problems, their behaviors must be resilient against such adversarial assaults.

Promisingly, there is recent evidence showing deep RL agents learn policies robust to adversary attacks at test time when they train with adversaries during learning \cite{Behzadan2017aa}. This has important implications for robust deep RL, as it suggests that security against attacks can be derived from learning. Here, we build on this idea and suggest that deriving adversary-aware agents from learning is a subset of the multi-agent reinforcement learning (MARL) problem.

At the heart of this problem is the need for an individual agent to coordinate its actions with those taken by other agents \cite{Fulda2007aa}. Given the role of inter-agent coordination in MARL, we suggest that operationalizing coordination between agent actions as a direct learning objective may lead to better policies for multi-agent tasks. Here, we present a quantitative metric that can be used to measure the degree of coordination between agents over the course of learning. Further, we present a research concept for using this metric to shape agent learning towards coordinated behavior, as well as the impact that different degrees of coordination can have on multi-agent task performance.

\subsection{Adversary-aware RL as MARL}
Understanding adversary-aware RL agents in terms of MARL is straightforward when we consider that training in the presence of adversarial attacks is similar to training in the presence of agents pursuing competing goals. In competitive RL, outcomes are often considered zero-sum, when agents reward/loss are in direct opposition \cite{Busoniu2008aa,Crandall2011aa}. In the case of attacks on RL agents, the adversary's goal is typically to learn a cost function that, when optimized, minimizes the returns of the attacked agent \cite{Pattanaik2017aa}. Thus, the adversary's reward is the opposite of the attacked agent's.

If we take seriously these comparisons, the problem of creating adversary-aware agents is largely one of developing agents that can learn to coordinate their behaviors effectively with the actions of an adversary so as to minimize the impact of its attacks. Thus, adversary-aware RL is an inherently multi-agent problem.

\subsection{Coordination in MARL}
In MARL problems, the simultaneous actions of multiple actors obfuscate the ground truth from any individual agent. This uncertainty about the state of the world is primarily studied in terms of 1) partial-observability wherein the information about a given state is only probabilistic \cite{Omidshafiei2017aa}, and 2) non-stationarity where the goal of the task is ``moving'' with respect to any individual agent's perspective \cite{Hernandez-Leal2017aa}.

To the extent that uncertainty from an agent's perspective can be resolved, performance in multi-agent tasks depends critically on the degree to which agents are able to coordinate their efforts \cite{Matignon2012aa}. With MARL collaborative goals, individual agents must learn policies that increase their own reward without diminishing the reward received by other agents. Simple tasks, such as matrix or climbing games, present straightforward constraints that promote the emergence of coordination between agents, as these small state-space problems make the pareto-optimal solution readily discoverable.

Matignon et al. \shortcite{Matignon2012aa} enumerate the challenges for cooperative MARL, and show that no single algorithm is successful at achieving better performance. Instead, existing algorithms tend to address specific challenges at the expense of others. Further, in more complex state-spaces pareto-optimal solutions can be ``shadowed'' by individually optimal solutions that constrain learned behavior to selfish policies \cite{Fulda2007aa}. This undermines the performance gains achievable through coordinated actions in MARL problems. For these reasons, coordination between agents can only be guaranteed in limited cases where the challenges of MARL can be reasonably constrained \cite{Lauer2000aa}. As such, partial-observability and non-stationarity are problems that must be overcome for coordination to emerge \cite{Matignon2012aa}. For complex tasks, modern advances with DNNs have leveraged joint action learning to overcome the inherent uncertainty of MARL \cite{Foerster2017aa}. Indeed these algorithms show improved performance over decentralized and independent learning alternatives.

Though this work is promising, we recently showed that when coordination is directly measured, it cannot explain the improved performance of these algorithms in all cases \cite{Barton2018aa}. Coordination between agents, as measured by the causal influence between agent actions (method described below), was found to be almost indistinguishable from hard-coded agents forced to act independently. This leads to an interesting question about how to achieve coordinated actions between learning agents in real-world tasks where there is strong interest for the deployment of RL-equipped agents.

\section{Approach}
A possible solution to overcome the challenges of MARL is to address coordination directly. This concept was recently put to the test in several simple competitive tasks being performed by two deep RL agents \cite{Foerster2018aa}. The study explicitly took into account an opponent's change in learning parameters during its own learning step. Accounting for opponent behavior during learning in this manner was shown to yield human-like cooperative behaviors previously unobserved in MARL agents.

In a similar thrust, we propose here that coordination should not be left to emerge from the constraints on the multi-agent task, but instead be a direct objective of learning. This may be accomplished by providing a coordination measure in the loss of a MARL agent's optimization step.

\subsection{A novel measure for coordination in MARL}
The first step towards optimizing coordinated behavior in MARL is to define an adequate measure of coordination. Historically, coordinated behavior has been evaluated by agent performance in tasks where cooperation is explicitly required \cite{Lauer2000aa}. As we showed previously, performance alone is insufficient for evaluating coordination in more complex cases, and does not provide any new information during learning. 

Fortunately a metric borrowed from ecological research has shown promise as a quantitative measure of inter-agent coordination, independent of performance. Convergent cross mapping (CCM) quantifies the unique causal influence one time-series has on another \cite{Sugihara2012aa}. This is accomplished by embedding each time-series in its own high dimensional attractor space, and then using the embedded data of one time-series as a model for the other. Each model's accuracy is taken as a measure of the causal influence between the two time-series. 

In multi-agent tasks, we can define collaboration to be the amount of causal influence between time-series of agent actions, as measured by CCM. The advantage of this metric is that it provides a measure of coordination between agents that is independent of performance, and thus can be used as a novel training signal to optimize coordinated behavior. Thus, coordination is no longer exclusively an emergent property of the task, but rather a signal for driving agents' learned behavior.

\subsection{Coordination in an example MARL task}

We propose an experimental paradigm that is designed to measure the role of coordination in a continuous cooperative/competitive task: online learning of coordination during multi-agent predator-prey pursuit. In this exemplary experiment, CCM is used as a direct learning signal that influences how agents learn to complete a cooperative task.

The task is an adaptation of discrete predator-prey pursuit \cite{Benda1985aa} into a continuous bounded 2D particle environment with three identical predator agents and a single prey agent. Predators score points each time they make contact with the prey, while the prey's points are decremented if contacted by any predator.

Typically, agent learning would be driven solely by the environmental reward (in this case, agent score). With this typical framework, coordination may emerge, but is not guaranteed (see \cite{Barton2018aa}). In contrast, CCM provides a direct measure of inter-agent coordination, which can be used to modify agent learning through the incorporation of CCM as a term in learning loss. This can be done either indirectly as a secondary reward or directly as a term applied during back-propagation. Thus, learned behavior is shaped by both, task success and inter-agent coordination.

This paradigm provides an opportunity for coordination to be manipulated experimentally by setting a desired coordination threshold. As agents learn, they should coordinate their behaviors with their partners and/or adversaries up to this threshold. Minimizing this threshold should yield agents that optimize the task at the expense of a partner, while maximizing this threshold would likely produce high dimensional oscillations between agent actions that ignore task demands. Effective coordination likely lies between these extremes. Thus, we can directly observe the impact of coordinated behaviors in a MARL environment by varying this coordination threshold. To our knowledge, this has not been previously attempted.

\section{Implications and Discussion}

Explicit coordination between agents can lead to greater success in  multi-agent systems. Our concept provides a paradigm shift towards making coordination between agents an intended goal of learning. In contrast, many previous MARL approaches assume that coordination will emerge as performance is optimized. In summary, we suggest that coordination is better thought of as a necessary driver of learning, as important as (or possibly more important than) performance measures alone.

Our proposed use of CCM as a signal for inter-agent coordination provides a new source of information for learning agents that can be integrated into a compound loss function during learning. This would allow agents to learn coordinated behaviors explicitly, rather than gambling on agents discovering coordinated policies during exploration.

With the addition of coordination driven learning, the policies an agent learns will not take into account adversary behavior by chance, but rather by design. Such an algorithm would actively seek out policies that account for the actions of partners and competitors, limiting the policy search space to those that reason over the behavior of other agents in the system. We believe this is a reasonable avenue for more efficiently training mulit-agent policies.

Driving learning with coordination creates an opportunity for the development of agents that are inherently determined to coordinate their actions with a human partner. This is important, as without such a drive it is not clear how to guarantee that humans and agents will work well together. In particular, if modeling of human policies is too difficult for agents, they may settle on policies that try to minimize the degree of coordination in an attempt to recover some selfishly optimal behavior. Forcing coordination to be optimized during learning ensures that agents only seek out policies that are well integrated with the actions of their partners.

Our concept, as presented here, is to promote coordinated behaviors in intelligent learning agents by providing a quantitative measure of coordination that can be optimized during learning. The importance of implementing coordination to overcome adversarial attacks in the MARL problem cannot be understated. Furthermore, an explicit drive towards coordinated behavior between intelligent agents constitutes a significant advancement within the fields of artificial intelligence and computational learning.

\subsubsection{Acknowledgements}
This research was sponsored by the Army
Research Laboratory and was accomplished under Cooperative Agreement Number
W911NF-18-2-0058. The views and conclusions contained in this document are those
of the authors and should not be interpreted as representing the official policies, either
expressed or implied, of the Army Research Laboratory or the U.S. Government. The
U.S. Government is authorized to reproduce and distribute reprints for Government
purposes notwithstanding any copyright notation herein.

\bibliography{bib.bib}
 \bibliographystyle{alec18}
\end{document}